\documentclass[prl,10pt,twocolumn,preprintnumbers,amsmath,amssymb,nofootinbib]{revtex4}
\usepackage{amsmath}
\usepackage{amssymb}
\usepackage{amsfonts}
\usepackage{eucal}
\usepackage[active]{srcltx}%
\usepackage{color}
\usepackage[all]{xy}
\usepackage{epsfig,graphicx,bm}

\newcommand{\ba}{\begin{array}}
\newcommand{\ea}{\end{array}}
\newcommand{\be}{\begin{equation}}
\newcommand{\ee}{\end{equation}}
\newcommand{\bea}{\begin{eqnarray}}
\newcommand{\eea}{\end{eqnarray}}
\newcommand{\bse}{\begin{subequations}}
\newcommand{\ese}{\end{subequations}}
\newcommand{\bi}{\begin{itemize}}
\newcommand{\ei}{\end{itemize}}

\definecolor{darkgreen}{rgb}{0,0.3,0}
\definecolor{darkblue}{rgb}{0,0,0.3}
\definecolor{darkred}{rgb}{0.7,0,0}

\begin{document}

\vspace*{2mm}
\title{Excitation Entanglement Entropy in 2d Conformal Field Theories}

\author{M.M. Sheikh-Jabbari$^\natural$, 
and  H. Yavartanoo$^\dag$}
\bigskip\medskip
\affiliation{$^\natural$    \textit{School of Physics, Institute for Research in Fundamental
Sciences (IPM), P.O.Box 19395-5531, Tehran, Iran}   \\
\smallskip
$^\dag$ {State Key Laboratory of Theoretical Physics, Institute of Theoretical Physics,
Chinese Academy of Sciences, Beijing 100190, China}}
\vfil

\setcounter{footnote}{0}

\begin{abstract}
\noindent

We analyze how excitations affect the entanglement entropy for an arbitrary entangling interval in a 2d conformal field theory (CFT) using the  holographic entanglement entropy techniques as well as  direct CFT computations. We introduce the \emph{excitation entanglement entropy} $\Delta_h S$,  the difference between the entanglement entropy generic excitations and their arbitrary descendants denoted by $h$. The excitation entanglement entropy, unlike the entanglement entropy,  is a finite quantity (independent of the cutoff), and hence a good physical observable.  We show that the excitation entanglement entropy is governed by a second order differential equation sourced by the one point function of the energy momentum tensor computed in the excited background state. We analyze low and high temperature behavior of the excitation entanglement entropy and show that $\Delta_h S$ grows as function of temperature. We prove an ``integrated positivity'' for the excitation entanglement entropy, that although $\Delta_h S$ can be positive or negative, its average value is always positive. 
We also discuss the mutual and multipartite information with generic excitations. 

\end{abstract}
\date{\today}
\maketitle

Entanglement is one of the specific characteristics of quantum systems. From a theoretical perspective, given a system described by a normalized  density matrix $\rho$ ($\text{Tr}\rho=1$),  entanglement can be quantified in various different measures built from $\rho$. Entanglement entropy (EE) is the simplest of such measures and is defined as the von Neumann entropy of the system
\be\label{EE-def}
S_{EE}=-\text{Tr} \rho \ln\rho.
\ee
Entanglement entropy has emerged as a valuable tool for probing quantum systems in diverse situations, from condensed matter physics \cite{Cardy-Calabrese-review, Calabrese-Cardy-I, Casini-2009} to quantum gravity \cite{EE-gravity,Solodukhin-review, VanRaamsdonk}. Nonetheless,  calculation of EE for a general quantum mechanical system remains a daunting task. 

In (local) quantum field theory settings there is a one-to-one map between the Hilbert space of the theory and field configurations on a constant time slice. One may ask for entanglement between any two regions A and B on a constant time slice separated by a region $\partial$A. The entanglement would then depend on the shape of the regions A and B, as well as the field profile in these regions. The latter may be viewed as excitations on the vacuum field configuration. In principle one may try to compute \eqref{EE-def} using standard path integral techniques, perturbatively in the coupling  \cite{Calabrese-Cardy-I, Casini-2009}. These, except for free field theories with simple shapes for $\partial$A, are formidable computations. 

However, in two dimensional conformal field theories (2d CFT's), where the entangling region A is a one dimensional interval, invoking the infinite dimensional conformal symmetry simplifies the calculations and EE can be computed analytically 
$S=\frac{c}{3}\ln \frac{L}{\epsilon}$, where $L$ is the length of entangling region, $\epsilon$ is the cutoff regulating the ultraviolet divergences and $c$ is the central charge  \cite{Calabrese-Cardy-II, HLW}. This is conveniently computed by employing the replica trick \cite{Calabrese-Cardy-I, Calabrese-Cardy-II}, using 
\be\label{Replica}
S_A=\lim_{n\to 1}\frac{1}{1-n}\ln {\mathrm Tr}\rho^n_{_A}.
\ee 

Computation of the entanglement entropy in 2d CFT's for a given entangling interval may be extended to cases where we have excitations in the  interval. These excitations may be caused by presence of a primary operator or descendants of primaries and the field theory could be at zero or non-zero temperature or angular (chemical) potential \cite{Alcaraz:2011tn, Temperature-excited-EE, 2d-CFT-excitation, Nozaki:2014hna, Caputa:2014vaa,Cardy:2014jwa}.
Cases where we have time dependent excitations (e.g. quenches) has also been studied and time dependence and dynamics of EE has been analyzed \cite{Roberts-2012, EE-time-evolution, EE-excited-time-evolution-1, Nozaki:2013wia, Asplund:2014coa}. In this Letter we analyze further the computation of EE for generic excited states in 2d CFT's.
\vskip 2mm

The AdS/CFT duality \cite{AdS-CFT-review} (or more practical versions of it, the gauge/gravity correspondence) may be used as a tool to compute EE, especially for strongly coupled field theories. In general there is the  proposal by Ryu-Takayanagi (RT) for computing the EE holographically using the AdS/CFT techniques \cite{RT-proposal, NRT-review}.  According to the RT proposal computation of EE is reduced to calculating the area of a surface which has the minimum area among surfaces attached to the boundary of the asymptotically AdS bulk  at the boundary of entangling region $\partial A$. The (holographic) entanglement entropy is given in terms of the area of this minimal surface $m(\partial A)$:
\be\label{HEE-RT-def}
S_{HEE}=\frac{Area(m(\partial A))}{4G_N},
\ee
where $G_N$ is the Newton constant. The RT proposal which was initially made for static gravity backgrounds have been extended to stationary cases \cite{HRT-proposal}.

The RT proposal for computing HEE becomes particularly simple for AdS$_3$ geometry \cite{RT-proposal, NRT-review}: the minimal surface, which is a one-dimensional curve in this case, becomes the geodesic connecting the two spacelike separated points at the AdS$_3$ boundary. The two points at the boundary are the endpoints of the entangling interval in the dual 2d CFT. This geodesic length is logarithmically divergent, due to the near boundary behavior of the metric. One may then regulate this length by a cutoff scale $\epsilon$. If the entangling interval at the boundary has length  $L$, then
\be\label{HEE-AdS3}
S_{HEE}=\frac{\mathrm{Length}_{geodesic}}{4G_3}=\frac{\ell}{2G_3} \ln\frac{L}{\epsilon},
\ee
where $\ell$ is the AdS$_3$ radius and $G_3$ is the 3d Newton constant. If  the presumed dual 2d CFT is at the Brown-Henneaux central charge $c$ \cite{Brown-Henneaux}
\be\label{central-charge}
c=\frac{3\ell}{2G_3},
\ee
the above exactly matches the 2d CFT direct computations. The above analysis may be repeated for other known locally AdS$_3$ geometries \cite{HRT-proposal}, like BTZ black holes \cite{BTZ}, successfully reproducing the entanglement entropy of a 2d CFT at non-zero temperature \cite{Calabrese-Cardy-II}.

Here we analyze  holographic computation of the EE for general asymptotic AdS$_3$ geometry, the Ba\~nados geometries \cite{Banados}, 
associated with generic excited states in the dual 2d CFT. The pure AdS$_3$ Einstein gravity is described by
\be\label{Einstein-eq}
{\cal S}=\frac{1}{16\pi G_3}\int d^3x \sqrt{-g}(R+\frac{2}{\ell^2})\  \Longrightarrow  R_{\mu\nu}+\frac{2}{\ell^2}g_{\mu\nu}=0.
\ee 
All solutions to \eqref{Einstein-eq} are locally AdS$_3$ geometries and are fully specified by the falloff (near boundary) behavior of the solutions and possibly other topological data \cite{3d-gravity}.  The most general solution obeying Brown-Henneaux boundary conditions \cite{Brown-Henneaux}, are the Ba\~nados geometries \cite{Banados}
\be
\label{general3dmetricZ}
ds^2=\ell^2\frac{dz^2}{z^2}-(\frac{\ell^2}{z} dx^+- z L_-dx^- )(\frac{\ell^2 }{z} dx^--  z L_+dx^+),
\ee
where $x^\pm\in [0,2\pi]$ and $L_\pm=L_\pm(x^\pm)$  are two smooth and periodic, but otherwise arbitrary, functions.  In the coordinate system adopted here the causal boundary of these geometries is located at $z=0$ and is (a part of) a cylinder $\mathbb{R}\times S^1$ spanned by $x^\pm$, where $\mathbb{R}$ is the time like direction which is along $x^++x^-$ and $S^1$ is parametrized by $x^+-x^-$. The above family contain the well-known BTZ black hole solutions \cite{BTZ} as a special class associated with  constant and positive $L_{\pm}\equiv {\mathcal T}_\pm^2$.  The inner and outer horizons radii $\rho_{\pm}$ of the BTZ black hole are given by
\be
\rho_{\pm}=\ell({\mathcal T}_+\pm {\mathcal T}_-).
\ee

Ba\~nados geometries and their geometric properties have been studied in some detail \cite{Banados-geometries-I,Banados-geometries-II}, where it was shown that these geometries have generically two global (compact) $U(1)$ Killing vectors. Moreover, it was shown in  \cite{Banados-charges} (see also \cite{Banados-geometries-II}) that 
there are two class of conserved charges associated with the above geometries: the charges associated with the two $U(1)$'s, the $J_\pm$ charges, and the  $L_n, \bar L_n,\ n\in \mathbb{Z}$ charge which form two copies of  Virasoro algebra both at the  Brown-Henneaux central charge \eqref{central-charge}. 
Importantly,   $J_\pm$ commute with $L_n, \bar L_n$ \cite{Banados-charges}.  Moreover, Ba\~nados geometries are in one-to-one relation with representations of the left and right Virasoro groups, $Vir_+\otimes Vir_-$ \cite{Banados-geometries-II}. These representations are usually called  Virasoro coadjoint orbits \cite{Witten, Balog}. {The ``thermodynamic'' properties and quantities of the geometries \eqref{general3dmetricZ} are shared among all geometries associated with the same orbit \cite{Banados-geometries-II}.}

Appearance of  Virasoro algebra suggests presence of a 2d CFT dual with central charge $c$ \eqref{central-charge} to the gravity theory \eqref{Einstein-eq}. According to the standard AdS/CFT dictionary \cite{AdS-CFT-review}  the Ba\~nados geometries are then dual to excited states in the presumed dual CFT and that the expectation value of the left and right components of the energy momentum tensor of the 2d CFT $T(x^+),\ \bar T(x^-)$ are related to the functions $L_\pm$:
\be\label{T-L}
\langle T(x^+)\rangle =\frac{c}{6} L_+\,,\qquad \langle \bar T(x^-)\rangle =\frac{c}{6} L_-\,.
\ee
The special case of $L_+=L_-=0$ corresponds to massless BTZ, and the vacuum state of the presumed dual 2d CFT on the plane (if we ignore the periodicity in $x^\pm$). 

The primary operators which satisfy the unitarity bound, in the large $c$ limit where we can trust classical gravity description, correspond to geometries with constant $L_\pm$ with $L_\pm\geq -1/4$ and Ba\~nados geometries in  Virasoro coadjoint orbits with constant representative $L_\pm$  correspond to generic descendants of these primaries in the presumed dual 2d CFT \cite{Banados-charges, Banados-geometries-II}. Note that the relation between generic excitations and the geometry is not one-to-one. As \eqref{T-L} shows, all 2d CFT excitations with equal energy-momentum expectation values correspond to the same geometry. With the same token, any observable which is built from the geometry, like the HEE, would at most be able to encode information about the geometry, in the case of Ba\~nados metrics; that is the information in functions $L_\pm$.

{In this work we focus on generic excitations associated with constant representative orbits and study how the change in the excitations affects the entanglement entropy.}

\textbf{Holographic computation of the  Entanglement Entropy.} Consider a 2d CFT on a cylinder parametrized by $x^\pm$:
\be\label{2d-metric}
ds^2=-\ell^2 dx^+dx^-=\ell^2 (-dt^2+d\phi^2), \ \phi\in[0,2\pi],
\ee
and an entangling interval whose endpoints are two spacelike separated points given by $x_1^\pm$ and $x_2^\pm$. For holographic computation of the entanglement entropy we follow similar idea used in \cite{HRT-proposal, NRT-review, Hubeny:2013gta, Roberts-2012}. To calculate the HEE associated with this interval for a generic excited state the expectation value of energy momentum tensor of which is given by \eqref{T-L}, we note that all geometries of the form \eqref{general3dmetricZ} are locally AdS$_3$. This means that there should be a coordinate transformation which locally brings the metric \eqref{general3dmetricZ} to a similar metric with $L_\pm=0$. One may readily check that such a coordinate transformation is of the form \cite{footnote-roberts}
\be\label{coord-transf}
\begin{split}
x^+&\to y^+=\int \frac{dx^+}{\psi^2}-\frac{z^2\phi'}{\psi^2\phi(1-z^2/z_H^2)},\\
x^-&\to y^-=\int \frac{dx^-}{\phi^2}-\frac{z^2\psi'}{\phi^2\psi(1-z^2/z_H^2)},\\
z&\to w=\frac{z}{\psi\phi(1-z^2/z_H^2)},
\end{split}
\ee
where $\psi, \phi$ are solutions to the Hill's equation \cite{Banados-geometries-I, Banados-geometries-II}
\be\label{Hill-eq}
\psi''- L_+\psi=0,\quad \phi'' -L_- \phi=0.
\ee
Each of the above two equations have two solutions, which may be denoted by $\psi_1,\psi_2$ and $\phi_1, \phi_2$ and one may choose the normalization 
\be
\psi_1'\psi_2-\psi_2'\psi_1=1,\ \ \phi_1'\phi_2-\phi_2'\phi_1=1.
\ee
In the above transformations \eqref{coord-transf} there are four choices for $\psi$'s and $\phi$'s and four associated $z_H^2$'s:
\be
z_H^2=\frac{\psi_i}{\psi_i'}\frac{\phi_j}{\phi_j'},\quad i,j=1,2.
\ee 
The $z_H$'s are radii of the Killing horizons of the Ba\~nados geometry \cite{Banados-geometries-II}. As expected, the coordinate transformation \eqref{coord-transf} is not globally defined, does not respect periodicity of $x^\pm$. Moreover, it is defined on one side of the inner and outer horizons. 

To compute the HEE, we only need to know the length of geodesics anchored at the boundary at $x_1^\pm,\ x_2^\pm$, note that this length is invariant under local diffeomorphisms and that it only depends on the separation of the endpoints at the boundary (at $z=0$). Therefore, for the HEE it suffices to know the near boundary behavior of \eqref{coord-transf}. Such a coordinate transformation at  $z=0$ is like a conformal map at the boundary with \cite{footnote-H}
\be
x^+\to H_+(x^+)=\frac{\psi_1}{\psi_2},\quad x^-\to H_-(x^-)=\frac{\phi_1}{\phi_2}.
\ee
Under this map
\be\label{L-transform}
L_\pm(x^\pm)\to  L^H_\pm(x^\pm)=H'^2_\pm L_\pm(H_\pm)-{\cal S}[H_\pm; x^\pm],
\ee
where $S[H_\pm;x^\pm]$ is the Schwarz derivative
\be
{\cal S}[H;x]=\frac{H'''}{2H'}-\frac{3H''^2}{4H'^2},
\ee
and \eqref{2d-metric} is mapped to
\be\label{scaled-2dmetric}
ds^2= -\ell^2 H'_+ H'_- \ dx^+ dx^-.
\ee
Here primes denote derivative w.r.t. the argument.  One may then check that ${\cal S}[H_\pm;x^\pm]=-L_\pm(x^\pm)$
and hence \eqref{L-transform} yields, 
$L_+^H(H_+(x^+))=0,\ L^H_-(H_-(x^-))=0.
$ 

One can use \eqref{HEE-AdS3} to readily compute the HEE for generic Ba\~nados geometry
\be\label{Banados-HEE}
S_{HEE}(x_1,x_2)= \frac{c}{6}\ln\left(
\frac{{\cal L}_+(x_1^+,x_2^+){\cal L}_-(x_1^-,x_2^-)}{\epsilon^2 }\right),
\ee
where
\be\label{cal-L}\begin{split}
{\cal L}_+(x_1^+,x_2^+)&=\psi_1(x^+_1)\psi_2(x^+_2)-\psi_2(x^+_1)\psi_1(x^+_2),\\
{\cal L}_-(x_1^-,x_2^-)&=\phi_1(x^-_1)\phi_2(x^-_2)-\phi_2(x^-_1)\phi_1(x^-_2).
\end{split}
\ee
Recalling the fact that $H'_+=1/\psi_2^2, H'_-=1/\phi_2^2$, due to the conformal factor $H'_+H'_-$ in the metric \eqref{scaled-2dmetric} we have adjusted the cutoff $\epsilon$  with the mapping $x\to H(x)$. \cite{footnote}.
We also note that $S_{HEE}(x_1,x_2)=S_{HEE}(x_2,x_1)$ which readily follows from \eqref{cal-L}.

For the vacuum case $L_\pm=0$, $\psi_1=x^+, \psi_2=1$ (and similarly for $\phi$'s) and therefore, ${\cal L}=x_1-x_2$. As another example, let us consider constant positive $L_\pm={\mathcal T}_\pm^2$ corresponding to BTZ black hole. For this case
\be\label{BTZ-psi}
\begin{split}
\psi_1=\frac{1}{\sqrt{2{\mathcal T}_+}}e^{{\mathcal T}_+ x^+}&,\quad \psi_2=\frac{1}{\sqrt{2{\mathcal T}_+}}e^{-{\mathcal T}_+ x^+},\\
\phi_1=\frac{1}{\sqrt{2{\mathcal T}_-}}e^{{\mathcal T}_- x^-}&,\quad \phi_2=\frac{1}{\sqrt{2{\mathcal T}_-}}e^{-{\mathcal T}_- x^-}.
\end{split}
\ee
Then \eqref{Banados-HEE} simplifies to
\be\label{HEE-BTZ} 
S_{HEE}=\frac{c}{6}\ln \left(\frac{\sinh({\mathcal T}_+R_+)}{{\mathcal T}_+\;\epsilon}\cdot \frac{\sinh({\mathcal T}_-R_-)}{{\mathcal T}_-\;\epsilon}\right),
\ee
where 
\be\label{Rpm}
R_\pm=min(\Delta x^\pm, 2\pi-\Delta x^\pm),\quad \Delta x^\pm=|x_1^\pm-x_2^\pm|,
\ee 
is the length of the entangling region (in units of AdS$_3$ radius $\ell$). For an entangling region with the two endpoints of the same time coordinate $R_+=R_-=R$, $R$ being the length of entangling interval, this result of course matches that of \cite{HRT-proposal, NRT-review, Hubeny:2013gta} once we replace ${\mathcal T}_\pm$ with the inverse temperature $\beta_\pm=\pi/{\mathcal T}_\pm$. 

\vskip 2mm
\emph{\textbf{Excitation Entanglement Entropy.}} Descendants of any given primary operator are constructed from the primary by the action of a conformal transformations. In the terminology of Virasoro coadjoint orbits, i.e. primary operators correspond to constant representative orbits \cite{Balog, Banados-geometries-II}.  At the level of the 3d dual geometries at least in the large $c$ limit, as pointed out, primary operators in the 2d CFT correspond to geometries with constant $L_\pm$ and the descendants are related through coordinate transformation which at the AdS$_3$ boundary are given as \cite{Banados-charges, Banados-geometries-II}
\be\label{conformal-map-h}
x^\pm\to \tilde{x}^\pm=h^\pm(x^\pm)\,,\ h_\pm(x^\pm+2\pi)=h_\pm(x^\pm)+2\pi,
\ee
and $h'_\pm>0$. Under this map
$L_\pm$ transform as \eqref{L-transform} while $\psi_1,\psi_2\to \tilde{\psi}_1, \tilde{\psi}_2$ where
\be\label{conformal-map-psi}
\tilde{\psi}_1=\frac{1}{\sqrt{h_+'}}\psi_1(h_+(x^+)),\
\tilde{\psi}_2=\frac{1}{\sqrt{h_+'}}\psi_2(h_+(x^+))\,,
\ee 
and similarly for $\phi$'s. The descendants of primaries are hence specified by $L_\pm$ functions of the form
\be\label{L-primary-descendant}
L_\pm= h'^2_\pm {\cal T}_\pm^2 -{\cal S}[h_\pm;x^\pm].
\ee

One may now readily check how  HEE \eqref{Banados-HEE} changes under conformal transformations \eqref{conformal-map-h} and hence obtain the HEE for the excitation generated by conformal transformtion $h$,
\be\label{HEE-Orbit-change}\begin{split}
&S^{(h)}_{HEE}(x_1,x_2)= S_{HEE}(h(x_1),h(x_2))-\Delta S_{EE}\\
&\Delta S_{EE}=\frac{c}{12}\ln(h'_+(x^+_1)h'_+(x^+_2)h'_-(x^-_1)h'_-(x^-_2)).
\end{split}\ee
We then define the \emph{excitation entanglement entropy} 
\be\label{EEE}
\Delta_hS(x_1,x_2)\equiv S^{(h)}_{HEE}(x_1,x_2)-S_{HEE}({x_1,x_2}),
\ee
i.e. $\Delta_h S$ computes the difference between the entanglement entropy of a (primary) excitation and its conformal descendants. For a 2d CFT, from \eqref{HEE-Orbit-change}
\be\label{EEE-2dCFT}
\Delta_h S (x_1,x_2)=\frac{c}{6}\ln \left({\cal X}_+\cdot {\cal X}_-\right)
\ee
where
\be\label{cal-X}
{\cal X}_\pm=\frac{{\cal L}_\pm(h_\pm(1),h_\pm(2))}{{\cal L}_\pm(x_1^\pm,x_2^\pm) \sqrt{h_{\pm}'(1)h_{\pm}'(2)}},
\ee
where $X_\pm(\alpha)=X_\pm(x^\pm_\alpha)$. The $\Delta_h S$ encodes both the information from the background (i.e. $L_\pm(x^\pm)$) and the excitations $h^\pm$ functions.
We note that the cutoff dependence has been dropped from the excitation entanglement entropy \eqref{EEE-2dCFT}. Moreover, the functions $h^\pm$ and $L_\pm$ can in principle be extracted if we know $\Delta_h S(x_1,x_2)$ for all $x_1,x_2$. 

As an example one may compute for the BTZ black hole and its conformal descendants.  Using \eqref{cal-L}, \eqref{BTZ-psi} and \eqref{conformal-map-psi}, we obtain
\be\label{EEE-generic}
\Delta_hS^{\text{BTZ}}(x_1,x_2)=\frac{c}{6}\ln\left( {\cal X}^{\text{BTZ}}_+\cdot {\cal X}^{\text{BTZ}}_-\right),
\ee
with
\be\label{X-BTZ}
{\cal X}^{\text{BTZ}}_\pm=\frac{\sinh({\mathcal T}_\pm\sqrt{h_\pm'(1)h_\pm'(2)} R_{h_\pm}) }{\sqrt{h_\pm'(1)h_\pm'(2)}\sinh({\mathcal T}_\pm R_\pm) },
\ee 
where $R_\pm$ is given in \eqref{Rpm} and 
\be\label{Rh-pm}
\hspace*{-4mm} R_{h_\pm}=\frac{min(\Delta h_\pm, 2\pi-\Delta h_\pm)}{\sqrt{h_\pm'(1)h_\pm'(2)}} ,\ \Delta h_\pm\equiv {h_\pm(1)-h_\pm(2)},
\ee
where $\Delta h_\pm$ shows the angular distance of the two endpoints of the entangling interval while $R_{h_\pm}$ is its physical length. (Note that after the map \eqref{conformal-map-h} the metric of the 2d boundary cylinder gets a conformal factor and hence the length of the space-like geodesic connecting them is not simply  difference of the $h$ functions at two end points.) For the special case of vacuum (massless BTZ) descendants with ${\mathcal T}_\pm=0$, the above reduces to 
\be\label{X-vac-desc}
{\cal X}_\pm^{\text{vac.desc.}}=\frac{ R_{h_\pm}}{ R_\pm}.
\ee
This matches with the results of \cite{Roberts-2012}.

\vskip2mm
\textbf{Comparison to 2d CFT results.} The above holographic results may also be directly verified by 2d CFT computations. To this end, we use the standard analysis of \cite{Calabrese-Cardy-I, Calabrese-Cardy-II} and the replica trick. Although one may perform this analysis in a more general form, using a generic density matrix $\rho$  with a primary, thermal or non-thermal excitation, for illustrative  purposes and since it yields the same results, here we present the computation for a pure state excitation $\rho=|\Psi\rangle\langle\Psi|$ where $\Psi$ denotes a generic of 2d CFT. We compute $\text{Tr}\rho^n$ for the interval $A$ whose endpoints are denoted by coordinates $z_1, z_2$ (as usual we are considering the Euclidean 2d CFT on $C$-plane with complex coordinates $z$):
\be
\text{Tr}\rho_{_A}^{n}=\frac{\langle\Psi|\sigma_n(z_1)\tilde\sigma_n(z_2)|\Psi\rangle_{\Sigma_n}}{(\langle\Psi|\Psi\rangle)^n},
\ee
where $\sigma, \tilde\sigma$ are the twist operators and $\Sigma_n$ is the standard $n$-sheeted plane \cite{Calabrese-Cardy-II}. 

Any given state $|\Psi\rangle$ which is a part of a unitary representation (orbit) of the 2d conformal group and other states in the same family/orbit are related  by a unitary transformation $U$ generated by conformal transformations (descendant generating transformation): 
\be
|\Psi_{(h)}\rangle=U_h|\Psi\rangle 
\ee
The conformal transformation { $U_h$} is associated with the conformal map \eqref{conformal-map-h} $z\to h(z), \bar z\to \bar h$. (Note that we are considering Euclidean 2d CFT on the plane and to compare it with our holographic gravity analysis we should replace $z$ with $x^+$, $\bar z$ with $x^-$, $h_+$ with $h(z)$ and $h_-$ with $\bar h$.)

Then, as discussed in \cite{Roberts-2012, Mandal-2014}, 
\be
\hspace*{-5mm}\text{Tr}(\rho_{\text{{(h)}\ {A}}}^{n})=\frac{\langle\Psi|(U_h^\dagger)^n\sigma_n(z_1)U_h^n (U_h^\dagger)^n\tilde\sigma_n(z_2)U_h^n|\Psi\rangle_{\Sigma_n}}{(\langle\Psi|\Psi\rangle)^n},
\ee
where 
\be
U_h^\dagger \sigma_n(z) U_h=h'(z)^{nh_\sigma} \bar h'(\bar z)^{nh_\sigma} \sigma_n(h(z)),
\ee
and $h_\sigma=\frac{c}{12}(n-\frac{1}{n})$. Following the lines of \cite{Roberts-2012, Mandal-2014} and recalling \eqref{Replica}, above hence yields
\be\label{2d-CFT-excited-S}
S^{(h)}(z_1,z_2)=S(h(z_1), h(z_2))-\frac{c}{6}\ln |h'_1 h'_2|,
\ee
where $h'_i=h'(z_i)$. We stress that \eqref{2d-CFT-excited-S} may be used to relate the entanglement entropy of any excitation (on the right-hand-side) and that of its conformal descendants generated by $U_h$ (on the left-hand-side). \eqref{2d-CFT-excited-S} is of course the same result as \eqref{HEE-Orbit-change} and hence leads to the same expression for the excitation entanglement entropy. For the particular case of vacuum excitations this result is precisely the well-known result of \cite{HLW} (see also \cite{Roberts-2012}) and is exactly what we have computed in \eqref{EEE-generic} with \eqref{X-vac-desc}.
\vskip2mm
\textbf{Low and high temperature behavior.}
In real situations we usually do not have access to systems at zero temperature, while probing  entanglement at low temperatures and studying low temperature behavior of the entanglement could be possible. Such an analysis for 2d CFT with a mass gap has been presented in \cite{Cardy:2014jwa}. As discussed in \cite{Banados-geometries-II}, the BTZ geometries associated with \eqref{BTZ-psi} and their conformal descendants have the same left and right temperatures ${\cal T}_\pm$. A system at temperature $T$ is associated with $2/T=1/{\cal T}_++1/{\cal T}_-$. Therefore, low temperature system corresponds to ${\cal T}_-$ finite and ${\cal T}_+\Delta x\ll 1$, which is dual to near extremal BTZ black hole. The leading behavior of $\Delta_h S$ \eqref{X-BTZ} is
\be\label{Low-temp}
\hspace*{-0.3cm}\Delta_h S^{\text{low-temp.}}-\Delta_h S^{{T=0}}=\frac{c}{144} T^2 \big((\Delta h_+)^2-(\Delta x_+)^2\big), \ee
where  $\Delta_h S^{T=0}=\Delta_h S^{\text{BTZ}}$ at ${\cal T}_+=0$.
As we see the deviation from the zero temperature is positive (recall that $h'_\pm >0$) and starts from temperature-squared. That is, with the same excitation function $h$, increasing the temperature will increase the excitation entanglement entropy. This is in contrast with the behavior of the thermal entropy of 2d CFT's which is linear in temperature. It is also different than the behavior in 2d CFT's with mass gap \cite{Cardy:2014jwa}.

At high temperature one may again expand \eqref{X-BTZ} for large ${\cal T}_\pm$ to obtain
\be\label{high-temp}
\Delta_h S\sim \frac{c}{6}\left[{\cal T}_+ \big(\Delta h_+-\Delta x_+\big)+{\cal T}_- \big(\Delta h_--\Delta x_-\big)\right].
\ee
As expected, the above is positive, that is the descendants (excited states) have bigger entanglement entropy than their primary operator excitations. Moreover, we have a linear temperature dependence. This linear behavior is expected, as at high temperature  the entanglement entropy and the thermal entropy should show the same temperature dependence, e.g. see \cite{HRT-proposal}. 
 
\vskip2mm
\textbf{Thermal vs. excitation entanglement entropy.} In our analysis both $x^\pm$ and $h_\pm$ are parametrizing coordinates on a circle, therefore, $\Delta x_\pm, \Delta h_\pm \in [0,2\pi]$. As we have already used and discussed  in the expression for entanglement entropy like \eqref{HEE-BTZ} and \eqref{X-BTZ}  $R_\pm, R_{h_\pm}$ which are the physical length of the entangling interval 
should be used (see also \cite{Asplund:2014coa}).  This matches with the fact that with pure state excitations the entanglement entropy of a region $A$ and its complement are equal \cite{Araki-Lieb}. One can prove that in presence of thermal excitations, the difference between the entanglement entropy of the two ``complementary'' regions, the two entangling intervals which appear once we specify two points on the circle,  is the thermal entropy of the corresponding black hole background/non-zero temperature 2d CFT system \cite{HRT-proposal}. In our case one may see this in the $\Delta_h S$: Let us take $\Delta x_\pm=2\pi$. Then we learn that $\Delta h_\pm=2\pi$ (recall \eqref{conformal-map-h}). Therefore, for this case, recalling \eqref{EEE-2dCFT}, 
$\Delta_h S=-\frac{c}{6} \ln (h'_+ h'_-)$ and is independent of temperature ${\cal T}_\pm$. (Note that the ratio of the physical length of the two entangling intervals associated with excitations $h$ is $h'_+ h'_-$.) Recall that $\Delta_h S$ is the difference between the entanglement entropy of 
an {excitation} and the corresponding descendant and that, all of the these states are associated with the same temperature ${\cal T}_\pm$ which is an orbit invariant quantity \cite{Banados-geometries-II}. Note also that this $\Delta_h S$ can be positive or negative.
\vskip2mm
 
\textbf{Small excitations.}  For small excitations around a thermal background with temperatures ${\cal T}_\pm$, i.e. for $h^\pm(x)=x^\pm+\epsilon_\pm(x)$, up to first order in $\epsilon$ we get:
\be\label{small-excitation}\begin{split}
\Delta_\epsilon S=\frac{c}{6}\bigg[&\frac{{\cal T}_+\Delta x^+}{\tanh({\cal T}_+\Delta x^+)}\frac{\Delta \epsilon^+}{\Delta x^+} -\frac12(\epsilon_+'(1)+\epsilon_+'(2))\cr +& x^-\text{-sector}\bigg],
\end{split}\ee
where  $\epsilon(\alpha)=\epsilon(x_\alpha)$ and $\Delta \epsilon^+=\epsilon_+(2)-\epsilon_+(1)$. For high temperature regime of ${\cal T}\Delta x\gg 1$ this matches with \eqref{high-temp} and for  ${\cal T}\Delta x\ll 1$ with the excitations above the vacuum \eqref{X-vac-desc},
\be
\Delta_\epsilon S=\frac{c}{3}\text{Re}\bigg(\frac{\epsilon(z_1)-\epsilon(z_2)}{z_1-z_2}
-\frac12\epsilon'(z_1)-\frac12{\epsilon'(z_2)} \bigg).\nonumber
\ee
Note that for arbitrary $z_1,z_2$, $\Delta_\epsilon S$ is not necessarily positive definite, even 
when the expectation value of energy momentum tensors associated with the excitations  $\delta_\epsilon T=-\frac{c}{12}\epsilon_+''', \delta_\epsilon \bar T=-\frac{c}{12}\epsilon_-'''$ are positive.

\vskip2mm
\textbf{Arbitrary excitation for small entangling interval.}  Another interesting limit to study is the arbitrary excitation for small entangling interval, i.e. when $x_1^\pm=x^\pm, x_2^\pm=x^\pm+\delta x^\pm$. For this case \eqref{EEE-2dCFT} yields
\be\label{small-interval}
\hspace*{-0.3cm}{\delta \Delta_h S(x)}=\frac{c}{36}\left(\Delta_{h_+}{L}_+\cdot (\delta x^+)^2+ \Delta_{h_-} L_-\cdot (\delta x^-)^2\right),
\ee
where $\Delta_h L$ is variation of $L$ under conformal transformation,
\be\label{Delta-L}
\Delta_{h_{\pm}} L_{\pm}=(h_{\pm}'^{2}-1){\cal T}^2_\pm-{\cal S}[h_{\pm};x^{\pm}].
\ee 
Recalling \eqref{T-L}, one may rewrite \eqref{small-interval} as
\be\label{Delta-S-diff-eq}
\bigg(\partial_{x^\pm}^2 \Delta_h S(x,y)\bigg)_{y=x}=\frac{1}{3}\Delta_{h_\pm} \langle T_{\pm}(x)\rangle.
\ee
The above equation directly relates the variation in the entanglement entropy to the variation in the expectation value of the energy momentum tensor in the 2d CFT. Recalling the periodicity conditions on $\Delta_h S$ and $h$ and 
\be\label{Delta-S-properties}\begin{split}
\Delta_h S (x,y)&=\Delta_h S (y,x),\ \cr
\Delta_h S(x,y)|_{x=y}&=0,\ \bigg(\partial_{x^\pm}\Delta_h S(x,y)\bigg)_{x=y}=0,
\end{split}\ee
\eqref{Delta-S-diff-eq} together with \eqref{T-L} and \eqref{Delta-L} completely determines $\Delta_h S(x,y)$. In fact, the solution is obtained to be precisely the one in \eqref{X-BTZ}. Note that while \eqref{Delta-S-diff-eq} is written only in terms of  2d CFT quantities, it has a natural interpretation in terms of first law of entanglement and how it is  related to perturbation of Einstein equation in the bulk \cite{Temperature-excited-EE, Myers-Mark, Grav-thermodynamics}.

\vskip2mm
\textbf{Integrated positivity.} For generic $h$ and $L$, $\Delta_h L$ and hence ${\delta\Delta_h S(x)}$,  do not have a definite sign for generic $h$ and for all values of $x$. Nonetheless, for the constant representative orbits one can prove that \cite{Witten, Balog}
\be\label{Delta-L-integrated-positivity}
\int_0^{2\pi} dx^\pm \Delta_{h_{\pm}} L_\pm\geq 0,
\ee
and hence we have an \emph{integrated positivity} for $\delta \Delta_h S(x)$, i.e. the average value $\partial^2_x \Delta_h S(x,y)|_{y=x}$ is non-negative. 

Similarly, for generic entangling interval $\Delta x^\pm$, as one can see from explicit expressions \eqref{X-BTZ}, $\Delta_h S(x,x+\Delta x)$ do not have a definite sign. Nonetheless one can  prove  integrated positivity relation for generic case given in \eqref{EEE-generic} with \eqref{X-BTZ}:
\be
\label{positivity} 
\int_0^{2\pi} dx^\pm \Delta_h S(x, x+\Delta x)\geq 0.
\ee
We note that the above integral is basically giving the average of the excitation entanglement entropy which is potentially a good physical observables in backgrounds where  translation symmetry is broken by the excitations.   
\vskip2mm
\textbf{Multi-partite information and for arbitrary excitations.} Besides the entanglement entropy there are other information theoretic measures involving more than one entangling interval. The simplest such quantity defined between two entangling regions $A, B$ ($A,B$ may have overlapping regions or not) is the mutual information defined as \cite{Nielsen-Chuang}
\be
I(A:B)= S_A+S_B-S_{A\cup B}
\ee
where $S$ on the RHS is the entanglement entropy. In the regions $A,B$ we may have various excitations, e.g. excitations by an operator or by its conformal descendants.  Using \eqref{HEE-Orbit-change}, we find that 
\be\label{mutual-info-desc}
I(A:B)_{\text{des}}(x_1^\pm, x_2^\pm)=I(A:B)_{\text{pri.}}(h^\pm(x_1), h^\pm(x_2)).
\ee
The above shows how the mutual information with {excitations} and their descendants are related to each other. Note that in the mutual information the $\Delta S_{EE}$ \eqref{HEE-Orbit-change} drops out. 
Similar result and statement can also be made for multi-partite information with a similar result as in \eqref{mutual-info-desc}.

\vskip2mm
\textbf{Concluding remarks and discussion.}
We have studied, holographically and within 2d CFT, the entanglement entropy for a class of { excitations} and their descendants which can be represented through the expectation value of the energy momentum tensor \eqref{T-L}, at zero or finite temperature. Although we mainly considered $L_\pm={\cal T}_\pm^2\geq 0$ cases, all of our results are also true for $ -1/4\leq L_{\pm}\leq 0$, by taking ${\cal T}$ to be imaginary.  

In particular, by introducing excitation entanglement entropy $\Delta_h S$ we quantified how  the entanglement entropy changes as we turn on conformal excitations. $\Delta_h S$ is cutoff independent and finite, and can be a good physical measure for the system. Our results are compatible with and extends the results of \cite{Alcaraz:2011tn}. 

As discussed $\Delta_h S$ encodes all the information about the excitation function $h$ and the expectation value of the energy-momentum tensor of the excitation operator $L_\pm$. Therefore, one can in principle read off all the information about the geometry from $\Delta_h S$. To obtain our results we used the map \eqref{coord-transf}. This coordinate transformation maps the outside horizon region of the Ba\~nados geometries  \cite{Banados-geometries-II} to AdS$_3$ in Poincar\'e patch. So, we would expect to be able to reconstruct the outside horizon geometry using the $\Delta_h S$, in the same manner outlined in \cite{VanRaamsdonk}.

The excitation entanglement entropy $\Delta_h S$ is closely related to the geometry (and metric) of the kinematic space, the geometry of CFT intervals. In this regard \eqref{Delta-S-diff-eq} and \eqref{Delta-S-properties} may be viewed as equations for metric of the kinematic space and the integrated positivity \eqref{positivity} as a statement about positive norms on the kimentic space. Such analyses has been initiated in \cite{Czech:2015kbp}. See also \cite{Myers-deBoer}.

Relative entropy \cite{Relative-entropy} is another information theoretic measure which may be used to compare an excitation due to a primary operator and it descendants. It is interesting to compute and compare relative and our  excitation entanglement entropy for the cases we analyzed here.

One may study how local operations, e.g.  local projections measurements,  affect the excitation entanglement entropy. This question can be posed and answered within the 2d CFT or the dual AdS$_3$ gravity setting. 
 It would be interesting to extend the corresponding analysis \cite{Local-projection}  for generic excitations we discussed here. It would in particular be interesting to explore the meaning of \eqref{Delta-S-diff-eq} and the integrated positivity \eqref{positivity} in this context.

\vskip2mm
\begin{acknowledgements}

We would like to especially thank Joan Simon for his role in the development of our analysis and results and for long discussions on various stages of this work. We would also like to that Pawel Caputa for collaboration in early stages of this work and Jan de Boer and Tadashi Takayanagi for comments on the draft.
The work of M.M. Sh-J is supported in part by Allameh Tabatabaii Prize Grant of
Boniad Melli Nokhbegan of Iran, the SarAmadan grant of Iranian vice presidency in science and technology and the ICTP network project NET-68 and the ICTP Simons fellowship. We would like to thank ICTP, Trieste for the workshop on aspects of 3d gravity and the hospitality in the course of this project.

\end{acknowledgements}
\vskip2mm

\textbf{Appendix: Proof of integrated positivity \eqref{positivity}.}\\
Denoting $I^{\pm}(\Delta x) =\int_0^{2\pi} dx^\pm \Delta_h S(x, x+\Delta x)$, we have already seen that for small intervals $\Delta x^{\pm}\ll 1$, \eqref{Delta-L-integrated-positivity} yields $I_{\pm}\geq 0$. To prove this statement for arbitrary intervals $\Delta x^{\pm}$, it is enough to show that $\frac{dI^{\pm}}{d\Delta x} \geq 0$. From   \eqref{X-BTZ} and \eqref{EEE-generic} we get
\be
\frac{dI^{\pm}}{d\Delta x} = \int_0^{2\pi} \left[\frac{{\mathcal T}_{\pm} h'_{\pm}(x^\pm+\Delta x^\pm )}{\tanh {\mathcal T}_{\pm} \Delta h_{\pm}}-\frac{{\mathcal T}_{\pm}}{\tanh  {\mathcal T}_{\pm} \Delta x^{\pm}}\right] dx^{\pm}. \nonumber
\ee
To show that the right hand side of above equation is positive we show that
\be\label{inequality}
 \int_0^{2\pi} \frac{{\tanh} {\mathcal T}_{\pm} \Delta x^{\pm} }{ \tanh {\mathcal T}_{\pm} \Delta h_{\pm}} h'_{\pm}(x^\pm+\Delta x^\pm)\;dx^{\pm} \geq 2\pi.
\ee
Noting that functions $h'_{\pm}$ and $\Delta h_{\pm}$ are positive everywhere and that $\tanh x$ is positive and increasing for $x>0$, then 
\be
\int_0^{2\pi} \frac{ \tanh {\mathcal T}_{\pm} \Delta h_{\pm}}{ \mathrm{tanh}  {\mathcal T}_{\pm} \Delta x^{\pm}} \leq \int_0^{2\pi} \frac{  \Delta h_{\pm}}{ \Delta x^{\pm}}=2\pi.
\ee
To prove the above, we have used the fact that
$$
 \int_0^{2\pi} h'_{\pm} \;dx^{\pm}=\int_0^{2\pi} \frac{  \Delta h_{\pm}}{ \Delta x^{\pm}} = 2\pi,
$$
and that for any two positive $2\pi$-periodic functions $f, g$, 
$$
\int_0^{2\pi} \frac{f}{g} dx \geq 2\pi \frac{\int_0^{2\pi}f dx}{\int_0^{2\pi}g dx}.
$$


\begin{thebibliography}{99}
\bibitem{footnote-H}
There are in fact four choices for $H_\pm$, associated with the four possible choices in \eqref{coord-transf}. These are explicitly $H_+=-\psi_2/\psi_1$ or $H_+=\psi_1/\psi_2$ and similarly for $H_-$. It is straightforward to see that our results and analysis for (excitation) entanglement entropy is independent of these choices.
\bibitem{footnote}
{Although we are using the mapping to AdS$_3$ in Poincare patch, one could have directly performed the HRT proposal \cite{HRT-proposal} to obtain the same result as \eqref{Banados-HEE}.}
\end{thebibliography}
\end{document}